\documentclass[twocolumn,floatfix,superscriptaddress]{revtex4}
\usepackage{graphicx}
\usepackage{amsmath}
\usepackage{latexsym}
\usepackage{epstopdf}
\usepackage{amsmath}
\usepackage{amssymb,amsmath}
\usepackage[toc,page]{appendix}
\newcommand{\beq}{\begin{equation}}
\newcommand{\eeq}{\end{equation}}

\begin{document}

\title{Identification of Defective Two Dimensional Semiconductors by Multifractal Analysis: The Single-layer ${\rm MoS_2}$ Case Study}

\author{Reza Shidpour}
\affiliation{Department of New Technologies, Shahid Beheshti, Tehran, Iran}

\author{ S.M.S. Movahed }
\affiliation{Department of Physics, Shahid Beheshti University, Velenjak, Tehran 19839, Iran}

\vskip 1cm

\begin{abstract}

wo dimensional semiconductor such as single-layer transition metal dichalcogenides (SL-TMD) have attracted most attentions as an atomically thin layer semiconductor materials. Typically, lattice point defects (sulfur vacancy) created by physical/chemical method during growth stages, have disadvantages on electronic properties.  However, photoluminescence (PL) spectroscopy is conventionally used to characterize single-layer films but until now it has not been used to show the presence of defects or estimate their population due to overall similarity of general feature PL spectra. To find a feasible and robust method to determine the presence of point defects on single layer ${\rm MoS_2}$ without changing the experimental setup, Multifractal Detrended Fluctuation Analysis (MF-DFA) and Multifractal Detrended Moving Average Analysis (MF-DMA) are applied on the PL spectrum of single layer ${\rm MoS_2}$. We compare the scaling behavior of PL spectrum of pristine and defective single layer ${\rm MoS_2}$ determined by MF-DFA and MF-DMA. Our results reveal that PL spectrum has multifractal nature and different various population of point defects (sulfur vacancy) on single layer ${\rm MoS_2}$ change dramatically multifractality characteristics (Hurst, H\"older exponents) of photoluminescence spectrum. It is exhibited creating more lattice point leads to smaller fluctuations in luminescent light that it can help to design special defect structure for light emitted devices. The relative populations of point defects are almost elucidated without utilizing expensive characterization instruments such as scanning tunneling microscopy (STM) and high resolution transmission electron microscopy (HR-TEM).
\end{abstract}
\maketitle

\section{Introduction}

One of the most promising TMD's, monolayer ${\rm MoS_2}$, presents a wide range of applications such as liquid pollutant sensor \cite{lei08}, selective gas sensors \cite{he12,li12,per13}, catalysis \cite{yu14,thu99,jar07}, nano-lubricants \cite{mcc08}, lithium ion battery anodes \cite{xia10,chan11}, field-effect transistors (FETs) with high mobility and current On/Off  ratios \cite{amani13,radis11} and phototransistor \cite{zhang13} . Several growth methods have been developed to prepare thin-layer ${\rm MoS_2}$, including mechanical exfoliation \cite{li14}, ionic species intercalation \cite{cunn12}, intercalation assisted exfoliation \cite{fan15,sant94}, liquid/solution exfoliation \cite{colem11}, physical vapor deposition \cite{hel00}, hydrothermal synthesis \cite{ypeng01}, thermolysis of single precursor containing Mo and S \cite{putz00}, electrochemical synthesis \cite{li04} and chemical vapor deposition \cite{naj13}. Generally, all of these methods have some disadvantages. For example, atomically thin flakes of ${\rm MoS_2}$ exfoliated by mechanical cleavage show small size of the flakes and their thickness and shape were not controllable. One of most persistent problems is point defect (generally sulfur vacancy) created during doing synthesis. These point defects cause to deplete excellent electron mobility of single layer ${\rm MoS_2}$ and to restrict its application in electronics for its low quality \cite{yuan14}. So that, finding simple spectroscopic method to find point defect population seems to be vital for more applications. Although, photoluminescence spectroscopy (PLS) is typical tool to estimate layer number of ${\rm MoS_2}$, but the presence of S-vacancies cannot be determined by it. We want to propose a mathematical method to utilize PL spectrum of single-layer ${\rm MoS_2}$ to identify S-vacancy presence and even defect population.
    Multifractal detrended fluctuation analysis (MF-DFA) is developed on the basis of detrended fluctuation analysis (DFA) which is used to analyze the fractal scaling properties and long-range correlations in noisy signals \cite{peng95,podob08}. Since the inception of (MF-) DFA, it has been widely applied to analyze time series in fields such as geophysics \cite{love12}, river flow \citep{movahed08}, and economical time series \citep{stan00,podob09,movahed17} Compared with traditional multifractal analysis, such as that using the fixed-size box counting algorithms \cite{alber98} and R/S (rescaled range) analysis \cite{bass94}, MF-DFA can handle not only stationary time series but also non-stationary time series with unknown trend and noise \cite{kant02}. Due to some disadvantages in the internal algorithm of MF-DFA, a new version of analysis which is so-called Multifractal Detrended Moving Average Analysis (MF-DMA) has been proposed \cite{ser07,Gu.g,Shao.y}. Nowadays, it has been clarified that a remarkably wide variety of natural systems can be characterized by long-range power-law cross-correlation behavior \cite{kant02,zhou08,hajian10}. Existence and determination of power-law cross-correlations would help to promote our understanding of the corresponding dynamics and their future evolutions.
    In this paper, we utilize multifractal detrended fluctuation analysis in order to determine point defects populations in single layer (SL) Graphene-like material. Our study includes following novelties and advantages:  For the first time, we apply multifractal analysis methods for characterizing photoluminescence spectra captured from experiments. The contribution of defects on photoluminescence spectra can not be recognized by common methods, while according to our robust approach, we are able to find out fingerprint of defects. We also demonstrate that relative populations of corresponding defects are elucidated using a typical PL spectrum without utilizing expensive characterization instruments such as scanning tunneling microscopy (STM) and high resolution transmission electron microscopy (HR-TEM). 
       The procedure of experiment is carried out as follows: SL-${\rm MoS_2}$ without point defects such as sulfur vacancies is fabricated and then we produce three types defective SL-${\rm MoS_2}$ by ion sputtering. The recorded data from PL spectrum of these samples is used as input for MF-DFA algorithm. Relevant parameters such as Hurst exponent and width of singularity spectrum reveal that all these PL spectra of both pristine and defective single layer ${\rm MoS_2}$ have multifractal nature. Also, multifractality characteristics of mentioned samples would be changed by presence of point defects population, consequently, we can discriminate pristine SL-${\rm MoS_2}$ among synthesized samples without any further spectroscopy/microscopy. Not only PL spectra can determine to be monolayer but also lattice point defects are identified by using MF-DFA  and MF-DMA methods.
    The structure of this paper is as follows: In section 2, theoretical notion of robust algorithm will be introduced. Section 3 is devoted to experimental approach to construct SL-${\rm MoS_2}$. Results and discussions including SL-${\rm MoS_2}$ fabrication and multifractal analysis of PL spectra will be given in section 4. Concluding remarks will be expressed in section 5.

\section{Theoretical Framework}

Robust analysis of data in presence of trends and unknown noise is a challenging computational approach in order to infer reliable knowledge. Many methods have been introduced and developed with wide range of capabilities. Among, various methods, Multifractal Detrended Fluctuation Analysis (MF-DFA) with known performance from computational point of view has been considered, extensively \cite{kant02}. Detrending procedure done by MF-DFA has discontinuity nature, therefore Multifractal Detrended Moving Average method was proposed \cite{ser07,Gu.g,Shao.y}. In this section for the sake of clarity, we will explain necessity parts of MF-DFA method developed by \cite{kant02} and MF-DMA introduced by \cite{ser07,Gu.g,Shao.y}.

\subsection{Multifractal Detrended Fluctuation Analysis}

The multifractal detrended fluctuation analysis (MF-DFA) is a modified version of detrended fluctuation analysis. The fundamental idea behind of the method is to estimate and remove the local trend superimposed on the signal. Detrended series is then examined to unravel fractal properties, such as power law behavior. To facilitate our discussion, MF-DFA is first introduced in brief in this subsection. Given a data sets, the main steps for this method are as follows \cite{kant02}:\\
	
{\bf Step 1:} Imagine that a recorded regular data in an experiment is represented by $\{x_k\}$.  We should compute the so-called '{\it profile}' according to: 
\begin{eqnarray}
Y(i)\equiv\sum_{k=1}^i \left[x_k-\langle x\rangle\right], \quad \quad i=1,...,N
\end{eqnarray}  
here $\langle \rangle$ means the average of $x_k$ and $N$ is number of data point.\\

{\bf Step 2:} Now for each given $s$, the '{\it profile}' can be divided into $N_s\equiv {\rm int}(N/s)$ non-overlapping local windows with equal length $s$. Here "int" is a function which takes the integer part of a number. Since $N/s$ may not be an integer, there could remain a short part of the '{\it profile}' which belongs to no local windows. In order to utilize that part of the series, the same procedure can be repeated starting from the opposite end of the series. As a result, $2N_s$ local windows are obtained altogether.\\

{\bf Step 3:} For the $\nu$th of the $2N_s$ local windows, the variance can be determined as
\begin{eqnarray}
&&F^2(\nu,s)\equiv{1 \over s} \sum_{i=1}^{s}\{Y[(\nu-1) s+ i] - y_{\rm fit}(\nu,i)\}^2,
\label{fsdef1}
\end{eqnarray}
for $ \nu=1,...,N_s$ and
\begin{eqnarray}
F^2(\nu,s)\equiv{1 \over s} \sum_{i=1}^{s}\{Y[N-(\nu-N_s)s+ i] - y_{\rm fit}(\nu,i)\}^2,
\label{fsdef2}
\end{eqnarray}
for $ \nu=N_s+1,...,2N_s$. Here $y_{\rm fit}(\nu,i)$ is the fitting polynomial which is treated as the local trend in the $\nu$th local
window. Actually, detrending procedure in DFA method is carried out by the subtraction of the polynomial fitting function from the profile data sets in each segment. To this end, polynomial function in $\nu$th partition is determined according to associated profile data, $Y$. The chi-square function will be minimized to achieve best fit parameters of fitting polynomial. The trend of order $m$ can be diminished by $y_{\rm fit}$ of order $m$ which is called MF-DFA$m$ \cite{kant02,physa,kunhu}. Previous studies confirmed that common trends are eliminated by selecting linear fitting function. No trend means one should take a zeroth-order fitting function \cite{PRL00}. Here,  we determine the best linear fit of the profile, therefore, we consider linear fitting function, $y_{\rm fit}$, for profile data  corresponding to a constant function for original series.

{\bf Step 4:} We obtain the $q$th order fluctuation function as:
\begin{equation}\label{ed:fluc1}
F_q(s) \equiv\left( {1 \over 2N_s} \sum_{\nu=1}^{ 2N_s} \left[F^2(\nu,s)\right]^{q/2} \right)^{1/q} 
\end{equation}
where, in general, the index variable $q$ can take any real value except zero. Using L'Hopital's rule for $q=0$ we get:
\begin{equation}\label{ed:fluc0}
F_0(s)\equiv \exp\left( {1 \over  4N_s} \sum_{\nu=1}^{ 2N_s}\ln F^2(\nu,s)\right) 
\end{equation}
Various values of the parameter $q$ enable us to quantify the contribution of different values of fluctuation functions in Eqs. (\ref{ed:fluc1}) and (\ref{ed:fluc0}). Consequently, for negative values of $q$, small fluctuations have dominant contribution in summation, on the contrary, positive values of $q$ cause the larger value of fluctuation contributions. It turns out that for $q=2$ the common DFA is retrieved.\\

{\bf Step 5:} Determine the scaling behavior of  fluctuation function, $F_q(s)$,  by analyzing the log-log plots of
$F_q(s)$ versus $s$  for each value of $q$. It is interesting that how the generalized 
fluctuation functions $F_q(s)$'s depend on the scale $s$ for different values of $q$. If the following relationship:
\begin{equation}\label{lam}
F_{q}(s) \sim s^{h(q)}
\end{equation} 
can be established, then we can see that $\{x_k\}$ has scaling behavior. Here, $h(q)$ is called the generalized Hurst exponent, since $h(q=2)$ is equal to the well-known Hurst exponent, $H$, for stationary series, while  $H=h(q=2)-1$ for non-stationary series \cite{taqqu} (see the appendix of \cite{movahed08,sadeghsun}). For $q>0$, the large fluctuations play dominant role in Eq. (\ref{ed:fluc1}), therefore $h(q)$ describes the scaling behavior large fluctuations. On the other hand for negative values of  $q$, the small fluctuations have dominant contribution in Eq. (\ref{ed:fluc1}) \cite{kant02}. For monofractal time series which are characterized by a single exponent over all scales, $h(q)$ is independent of $q$, whereas for a multifractal time series, $h(q)$ varies with $q$. This dependence is considered to be a characteristic property of multifractal processes. According to the value of $H$, $\{x_k\}$ is anti-correlated (if $H<0.5$), uncorrelated (if $H=0.5$) or long-range correlated (if $H>0.5$). Thus, $H$ is a very useful index for unraveling long-range dependence of underlying series. 
 
 The $h(q)$ obtained from MF-DFA is related to the Renyi exponent $\tau(q)$ for a one-dimensional data by:
\begin{equation}\label{tauqq}
\tau(q)=qh(q)-1.
\end{equation}
Another way to characterize a multifractal series is the singularity spectrum, $f(\alpha)$ relating to $\tau(q)$ via a Legendre transform and it is defined by 
\begin{equation}
f(\alpha)=q\alpha-\tau(q)=qh(q)\left(\frac{d\ln h(q)}{d\ln q}+1\right)
\end{equation}
where $\alpha=d\tau(q)/dq$ is knows as H\"{o}lder exponent or singularity strength which characterizes the singularities in series. 
The singularity spectrum, $f(\alpha)$, describes the singularity content of the underlying datasets. For a typical multifractal series, we have a spectrum for 
$\alpha$ instead of having single value. The interval
of H\"{o}lder spectrum, $\alpha\in [\alpha^{\rm
min},\alpha^{\rm max}]$, reads as \cite{muzy94,muzy95,halsey86}:
\begin{eqnarray}\label{holder1}
\alpha^{\rm min}&=&\lim_{q \rightarrow +\infty} \frac{\partial \tau(q)}{\partial q},
\end{eqnarray}
\begin{eqnarray}\label{holder2}
\alpha^{\rm max}&=&\lim_{q \rightarrow -\infty} \frac{\partial \tau(q)}{\partial q}.
\end{eqnarray}
According to generalized Hurst exponent, generalized multifractal dimensions reads as:
\begin{equation}\label{eq:dq}
D(q)=\frac{\tau(q)}{q-1}
\end{equation}
%that are used instead of  in some papers \cite{payam06}. 

\subsubsection{Multifractal Detrended Moving Average Analysis}

As mentioned before, the fitting procedure is done in each non-overlapping segment yielding a discontinuity for computing fluctuation functions at the boundary of each partition. The MF-DMA method has been introduced to resolve this problem \cite{ser07,Gu.g,Shao.y}. The only difference between DFA and DMA is the partitioning and computing residuals function. These part are described as follows:

(2): After making profile for input data  denoted by $Y$, we compute  the moving average function $\widetilde{Y(j)}$:
\begin{equation}
\widetilde{Y(j)}=\frac{1}{s}\sum_{-\lfloor(s-1)\theta\rfloor}^{\lceil(s-1)(1-\theta)\rceil}Y(j-k),
\end{equation}
The symbol $\lfloor a \rfloor$ indicates the largest integer value not greater
than $a$ and $\lceil a\rceil$  corresponds to the smallest integer value
not smaller than $a$. Also $\theta$ is the position parameter varying from zero to unity. Therefore, the $\theta=0$ is called the backward moving average, $\theta=0.5$ refers to the centered moving average, while $\theta=1$ is associated with the forward moving average \cite{Gu.g,xu05}.

(3): Detrended data is constructed by subtracting computed moving average function from the cumulative series $Y$ as:
\begin{equation}
\varepsilon(i) \equiv Y(i)-\widetilde{Y(i)},
\end{equation}
where $s-s_1\leq i\leq N-s_1$

(4): Now,  we divide $\varepsilon(i)$ into $N_s= int[N/s]$ non-overlapping windows with the size of $s$ and then we have fluctuation function as follows:
\begin{equation}
F^2(\nu, s) = \frac{1}{s}\sum_{i=1}^{s}\varepsilon^2(i+({\nu}-1)s).
\end{equation}
The rest parts are similar to that of explained in MF-DFA algorithm.

\section{Experimental Method}
Firstly, these films were grown on a ${\rm SiO_2}$ substrate using a CVD method. Then the samples were initially characterized in air using Raman and PL spectroscopy at room temperature \cite{shid16}. Experiments were performed on films of single layer ${\rm MoS_2}$. The samples were also characterized in vacuum using PL spectroscopy. The experiment takes place in a vacuum chamber with a base pressure of $7\times10^{-10}$ Torr. A Varian sputter gun was used for generating ${\rm Ar^+}$ ions. The gun is operated at $600$ V acceleration potential, $30$ mA emission current and sputter time of 4 seconds.  The experiment consisted of cycles of sputtering in room temperature followed by in situ measurement of photoluminescence.  To avoid sample degradation, all measurements were conducted in succession to one another.
\begin{figure}
\begin{center}
\includegraphics[width=1\linewidth]{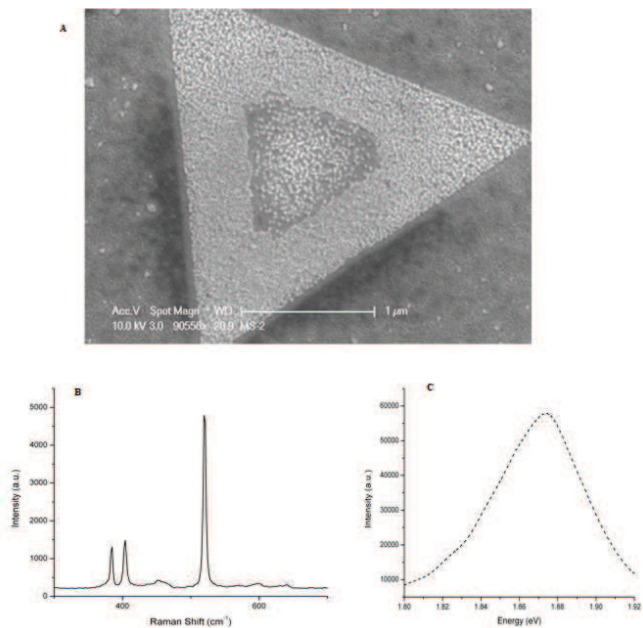}
\caption{\label{fig1} A) SEM image of ${\rm MoS_2}$ on ${\rm SiO_2/Si}$ substrate. Close-up image of a triangular ${\rm MoS_2}$ cluster after Pb sputtering. B) Raman spectra of ${\rm MoS_2}$ samples. The peak intensity ratio ${\rm E_{2g}/A_{1g}}$ is less than 1 for sample, confirming growth method leads to improved film quality. C) Spectra of single-layer ${\rm MoS_2}$. The PL intensities are normalized by the intensities of the ${\rm A_{1g}}$ Raman modes. The excitation wavelength is 532.5 nm.}
\end{center}
\end{figure}

\section{Results and discussion}
In this section we  will give our results after implementation of MF-DFA and MF-DMA , to evaluate multiscaling behavior of  the different samples ${\rm MoS_2}$.

\subsection{${\rm MoS_2}$ single-layer}

From the scanning electron microscopy (SEM) image shown in Fig. 1.A,  it is found that there were triangular ${\rm MoS_2}$ clusters in area about $1{\rm cm} \times 1{\rm cm}$ . The typical side length of the triangles was about $3\mu$m, which is also confirmed by a close up SEM image. Today, there is a well-defined and standard characterization method by photoluminescence and Raman spectroscopy to demonstrate single-layer and multilayer ${\rm MoS_2}$. In this method that has completely been confirmed by TEM and AFM measurements \cite{thu99,radis11}, two characteristics of Raman spectroscopy can be used to determine the number of layers of ${\rm MoS_2}$ samples. As reported, firstly, the position of ${\rm E_{2g}}$ and ${\rm A_{1g}}$ peaks and secondly, the frequency difference between the ${\rm E_{2g}}$ and ${\rm A_{1g}}$ that the former can be shifted a little depended to growth conditions but the latter is strongly related to layer number. The ${\rm E_{2g}}$ mode corresponds to in-plane vibrations (two S atoms are displaced in one direction and the Mo atom is displaced in the opposite direction). The ${\rm A_{1g}}$ mode corresponds to perpendicular vibrations (two S atoms are displaced in opposite directions while the Mo atom does not move). For single-layer ${\rm MoS_2}$, the peak position of ${\rm E_{2g}}$ is between 404 and $406 {\rm cm}^{-1}$ and ${\rm A_{1g}}$ is between 383 and $386 {\rm cm}^{-1}$ \cite{yu14,wang13,ling14}.  In addition, the difference between the ${\rm E_{2g}}$ and ${\rm A_{1g}}$ Raman modes which is widely used to identify the layer number 39 is about $20 {\rm cm}^{-1}$ for monolayer ${\rm MoS_2}$ , $~ 22 {\rm cm}^{-1}$ for bilayer, $~ 23 {\rm cm}^{-1}$ for trilayer and about $25 {\rm cm}^{-1}$ for bulk ${\rm MoS_2}$ \cite{yu14,yu13,chen13}.

As shown in Fig. \ref{fig1}.B, single-layer ${\rm MoS_2}$ sample exhibits two Raman characteristic bands for the ${\rm E_{2g}}$ mode at 384.85 ${\rm cm}^{-1}$ and ${\rm A_{1g}}$ mode at 404.08 ${\rm cm}^{-1}$ attributing to the single layer component.  The frequency difference for the single layer is 19.23 ${\rm cm}^{-1}$,which is identical to the exfoliated monolayer, 19.4 ${\rm cm}^{-1}$ unlike other CVD synthesized monolayer with previous results around 20.4 ${\rm cm}^{-1}$ \cite{chen13}. This could be related with certain crystalline imperfection, for example, smaller crystalline grains in the synthesized thin film. The full-width-half-maximum (FWHM) of peak can be used to determine the crystalline quality of the sample. In general, the smaller the width, the higher the crystalline quality of the sample \cite{aiz99}. The FWHM values related to the ${\rm E_{2g}}$ is 4.4 ${\rm cm}^{-1}$, close to that of the exfoliated monolayer, 3.7 ${\rm cm}^{-1}$. This suggests a good crystalline quality in the synthesized film. There is a direct 1.8-1.9 eV band gap equaled with 652-689 nm for single layer ${\rm MoS_2}$  that leads to strong photoluminescence (PL) \cite{zhu11}. This strong PL intensity depends to layer number and it only is considerable in monolayer and bilayer ${\rm MoS_2}$ so it can be used as an good indicator for mono layer/bilayer. In Fig. \ref{fig1}.C, the PL spectrum shows pronounced emission peak at 660 nm equalled to 1.88 eV with a superior photoluminescence ($\sim 50000$). 
\begin{figure}
\begin{center}
\includegraphics[width=1\linewidth]{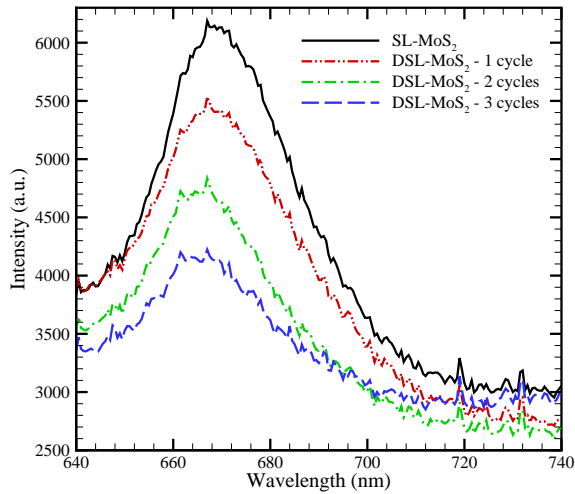}
\caption{\label{fig2} Photoluminescence spectrum of four samples from $620-800$nm. The excitation wavelength is $532.5$nm.  Solid line corresponds to single layer ${\rm MoS_2}$. Dashdotdot, dashdot and long-dash lines  are related to defective single layer ${\rm MoS_2}$ after 1 cycle, 2 cycles and 3 cycles, respectively. }
\end{center}
\end{figure}
\subsection{Defective single-layer ${\rm MoS_2}$}
    Once a sufficiently homogeneous area of the ${\rm MoS_2}$ film exhibiting exclusively single-layer Raman and PL characteristics had been identified, the sample was attached to a temperature-controlled manipulator in an ultra-high vacuum system. For subsequent studies of sputtering, the system was evacuated and baked to reach a base pressure of $10^{-7}$ Torr. A varian sputter gun operated at 600V acceleration potential, 30 mA emission current, and $3\times 10^{-5}$ and $7\times 10^{-6}$ Torr partial pressure of Ar was used for generating ${\rm Ar}^+$ ions. The sputter beam had a diameter of 0.3 cm. The PL experiments was done at a wavelength of 532 nm, and a liquid-nitrogen cooled Instruments detector. Our measurements involved cycles of sputtering at room temperature. To avoid sample degradation, all experiments were conducted in immediate succession to one another, with the sample maintained in ultra-high vacuum. %Within the duration of this experiment, we observe a reduction of the sulfur content of the film in the ${\rm MoS_2}$ structure. 
    To explore the impact of sputtering on the optical response of our films, we performed in situ PL measurements. Fig. \ref{fig2} shows PL spectra acquired at temperature of 300 K. Semi-conductor materials such as single layer ${\rm MoS_2}$ single layer have a gap in their band structure. With photon absorption by SL-${\rm MoS_2}$, the electrons excite from valence band to conduction band and this excitation leads to a general peak in photoluminescence spectrum of SL-${\rm MoS_2}$ as shown at Fig.  \ref{fig1}.C.  There are four samples including single layer ${\rm MoS_2}$ (SL-${\rm MoS_2}$), defective single layer ${\rm MoS_2}$ after 1 cycle ion sputtering (DSL-${\rm MoS_2}$-1 cycle), defective single layer ${\rm MoS_2}$ after 2 cycles ion sputtering (DSL-${\rm MoS_2}$-2 cycles), defective single layer ${\rm MoS_2}$ after 3 cycles ion sputtering (DSL-${\rm MoS_2}$ cycles).

\begin{figure}
\begin{center}
\includegraphics[width=1\linewidth]{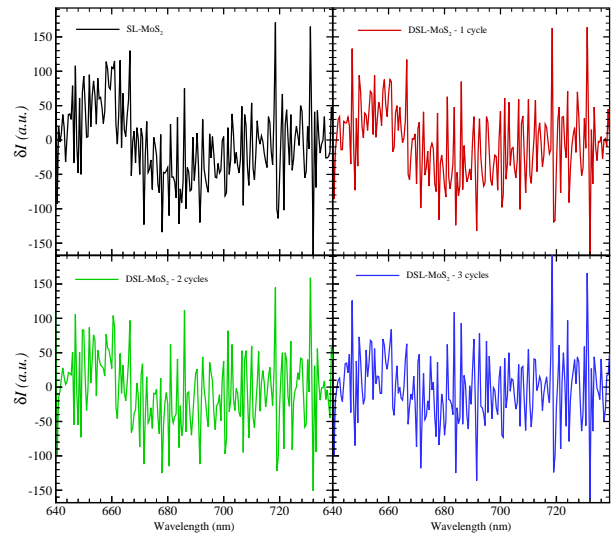}
\caption{\label{fig3} Increment data constructed from photoluminescence spectrum of samples.}
\end{center}
\end{figure}

When a defect is created in this material by ion sputtering or thermal annealing, an energy level was made below conduction band or upper valence band in forbidden Fermi zone. This means that photoluminescence spectrum would be noisy near general peak of it corresponding with electron hopping from valence band to conduction band.  If we only pick up the data near general peak, this noise in spectrum data near to general peak would be clearer as shown in Fig. \ref{fig2}.

\subsection{Implementation of MF-DFA and MF-DMA on PL Spectra of single layer ${\rm MoS_2}$}
The intrinsic point defects in the bulk and surface of semiconducting materials would introduce defect energy levels inside the band gap same as dopant atoms. The existence of defects in the chemical and structural composition of those materials can affect their optical and transport properties. A broad peak at $\sim 670$ nm ($\sim 1.8-1.9$ eV) in the optical spectrum of single layer ${\rm MoS_2}$ has been associated to impurities. In single layer ${\rm MoS_2}$, point defects are typically sulphur point defects imposed by elimination of sulfur atoms randomly distributed over the sample. Defects in the samples lead to the appearance of a series of peaks in the gapped region of the density of states (DOS). They are associated to the creation of midgap states localized around the defects, whose energy and strength depends on the specific missing atoms, their concentration, as well as the specific arrangement of the point defects as individual missing atoms \cite{yuan14}.  Our assumption is that effect of these peaks in gapped region of DOS can be reflected in PL spectrum around peak of spectrum. With this argument, only electron transition near Fermi level should be considered and consequently, the data in 640 - 740 nm interval is selected instead with 620 - 800 nm for further analysis. To apply MF-DFA or MF-DMA algorithms, we construct the increment data from original spectrum data as $\delta I(k)\equiv I(k+1)-I(k)$ (Fig. \ref{fig3}). 

\begin{figure}
\begin{center}
\includegraphics[width=1\linewidth]{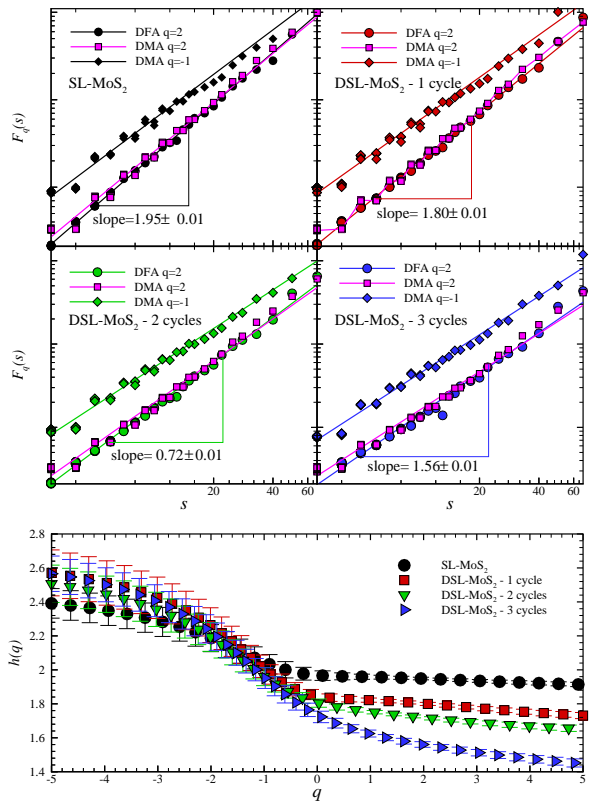}
\caption{\label{fig4} {\it Upper panel:}  Fluctuation function versus size of window,  $s$  in log-log plot. In this panel, filled circle symbols corresponds to DFA with $q=2$, filled square symbols is devoted to DMA with $q=2$ and filled diamond symbols is associated with DMA with $q=-1$. To make more sense we vertically shifted the $F_q(s)$.   {\it Lower panel:} Generalized Hurst exponent for all samples.}
\end{center}
\end{figure}

  We check the validity of applying this algorithm on our data by determining the scaling behavior of the fluctuation function, $F_q(s)$, versus the number of window, $s$, for various values $q$ according to Eqs. (\ref{ed:fluc1}) and (\ref{ed:fluc0}) or corresponding functions in MF-DMA method. The log-log plot of $F_q(s)$ versus $s$  (Fig. \ref{fig4}) is exhibiting the linear behavior in range of $s$ between 3 and 33 meaning that the scaling behavior exists for available data samples. To make more sense, we also utilize MF-DMA method on our data sets. Rectangular symbols in Fig. \ref{fig4} represents the scaling behavior of fluctuation function computed by MF-DMA method for $q=2$ and $\theta=0.5$, while circle symbols are devoted to the same function determined by MF-DFA. In addition in this figure diamond symbols illustrate the $F_{q=-1}(s)$ confirming the scaling behavior for $q=-1$. The solid lines in Fig. \ref{fig4} correspond to a power-law fitting function to data points.  The generalized Hurst exponents, $h(q)$, shows a $q$-dependency relationship (lower panel of Fig. \ref{fig4}). If  $h(q)$ changes with $q$  based on the nature of data under investigation, the underlying signal is considered to be multifractal. For monofractal series, $h(q)$ is constant.  Our results for other values of $q$'s  from MF-DMA and MF-DFA methods are consistent and hereafter we show only results determined by MF-DFA.   

As indicated in the lower panel of Fig. \ref{fig4}, the statistical errors of $h(q)$ with  $q<0$ for different samples have overlap, consequently, we consider the values of $h(q)$ with $q>0$ to compare four PL spectra. The $h(q)$ values of four samples for $q>0$ clearly demonstrate different behavior among single layer ${\rm MoS_2}$ thin films. The value of $h(q=2)$   determined by DFA is greater than one confirming our data is non-stationary. In this case the so-called Hurst exponent is given by $H=h(q=2)-1$. The value of Hurst exponent for all increment data extracted from PL spectrums is greater than 0.5 demonstrating that our data sets are classified in long-range correlated process. In addition, increasing point defects in SL-${\rm MoS_2}$ causes that the value of Hurst exponent approaches to 0.5 proceeding series to un-correlated behavior.  This means that the slope of fluctuation functions for $q>0$ computed for profile data decreases by increasing defects.  These fluctuations come from coupling vibrational lattice modes (phonons) with emission-induced electron transition \cite{alk12}. These vibrations that couple strongly to the distortion of atomic the geometry are expected to be dominant in the atoms close to the defect.
   
\begin{figure}
\begin{center}
\includegraphics[width=1\linewidth]{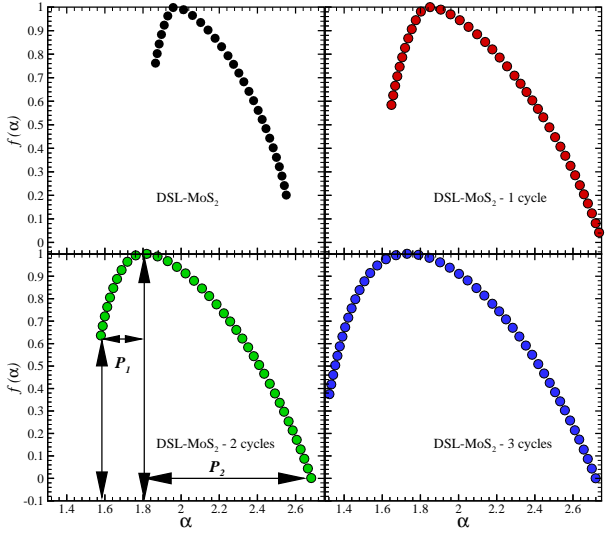}
\caption{\label{fig5} Singularity spectrum, $f(\alpha)$  as a function of H\"older exponent  of SL-${\rm MoS_2}$, DSL-${\rm MoS_2}$-1 cycle, DSL-${\rm MoS_2}$-2 cycles  and DSL-${\rm MoS_2}$-3 cycles.}
\end{center}
\end{figure}

   Fig. \ref{fig5} illustrates the multifractal singularity spectrum  which is another measure to quantify multifractal behavior of datasets. The width of singularity spectrum ($\Delta \alpha\equiv \alpha_{\rm max}-\alpha_{\rm min}$) determines the degree of multifractality of underlying dataset. The broader singularity spectrum the stronger multifractal nature of the signal.  The $f(\alpha)$ vs. $\alpha$ plot, depicting the multifractal singularity spectra (Fig. \ref{fig5}) provides a clear information about the relative presence of small and large fluctuations in the data of PL spectrum. With creating more point defects in SL-${\rm MoS_2}$, the ($\alpha_{\rm max}-\alpha_{\rm min}$) which indicates the broadening of spectrum increase  and consequently, the system goes to stronger multifractality. In multifractal singularity spectra, the position of $f_{\rm max}(\alpha)$  is important. If a large portion of the spectrum is tilted to the right $(P_1<P_2)$, it means that the spectrum is dominated by smaller (finer) fluctuations. On the contrary, higher value of large fluctuation represents $P_1>P_2$  with respect to mean value (see Fig. \ref{fig5}). Such singularity spectra shape has direct relevance to the shape of Hurst exponent plots $h(q)$, such that, the Hurst exponents corresponding to positive $q$ correspond to the left portion of the spectra with respect to $f_{\rm max}(\alpha)$. It turns out that for monofractal data, $f(\alpha)=1$ for $\alpha=h(q=2)$ and $h={\rm constant}$. A careful look to the multifractal spectra illustrated in Fig. \ref{fig5} confirms that there is a shift in the position of $f_{\rm max}(\alpha)$ towards lower values of $\alpha$ by increasing defect population. In other hand, presence of more point defect population leads to smaller fluctuations in increment PL spectrum data due to electron-phonon coupling.  We report the values of Hurst exponent and width of singularity spectrum ($\Delta \alpha$) determined by MF-DFA method in Table 1 at $1\sigma$ confidence interval.

\begin{table}
\begin{center}
%\medskip
\scalebox{0.7}{
\begin{tabular}{|c|c|c|}
  \hline
   Sample & $H$& $\Delta \alpha$  \\\hline 
  SL-${\rm MoS_2}$ &$0.95\pm0.01$& $0.69\pm0.05$ \\\hline
  DSL-${\rm MoS_2}$-1 cycle &$0.80\pm0.01$&$1.13\pm0.05$ \\\hline
  DSL-${\rm MoS_2}$-2 cycles &$0.72\pm0.01$&$1.14\pm 0.05$ \\\hline
DSL-${\rm MoS_2}$-3 cycles &$0.56\pm0.01$&$1.35\pm0.05$ \\\hline
\end{tabular}}
\caption{\label{tab1}The values of some relevant multifractal parameters at $1\sigma$ confidence interval.}
\end{center}
%\end{widetext}
\end{table}

\begin{figure}
\begin{center}
\includegraphics[width=1\linewidth]{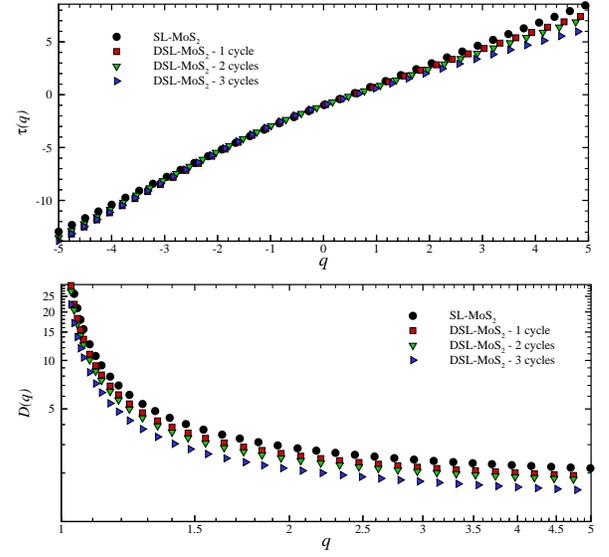}
\caption{\label{fig6} {\it Upper panel:} Multifractal scaling exponent, $\tau(q)$, for each sample. {\it Lower panel:} Generalized multifractal dimension, $D(q)$, for our samples.}
\end{center}
\end{figure}

Other function that can distinguish among the samples spectra is Multifractal scaling exponent, $\tau(q)=qh(q)-1$, that reveals the multifractality in data with various slope of  $\tau(q)$ for different $q$. As shown at upper panel of Fig. \ref{fig6}, when the population of defects on SL-${\rm MoS_2}$ increases, the slope of $\tau(q)$ decreases for $q>0$ from $1.906$ to $1.717$, $1.633$, and $1.415$, respectively. Generalized multifractal dimension defined by Eq. (\ref{eq:dq}) for PL SL-${\rm MoS_2}$ is shown in lower panel of Fig. \ref{fig6}. 

\begin{figure}
\begin{center}
\includegraphics[width=1\linewidth]{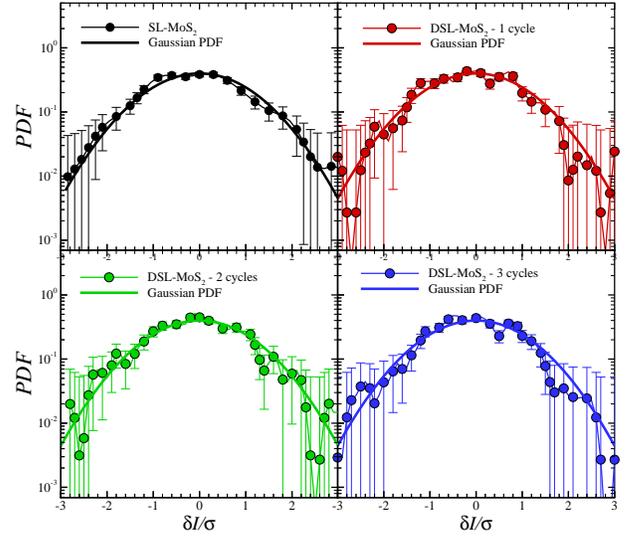}
\caption{\label{figpdf} Probability density function for our sample data. The symbols corresponds to direct numerical computation while the solid lines in each part indicates a Gaussian PDF. Here we rescaled data sets to unit variance, $\sigma$. }
\end{center}
\end{figure}

As explain in more details in various researches \cite{kant02}, in principle two different types of multifractality in time series can be distinguished. The broadness (fatness) of probability density function (PDF) and/or the different long-range correlations of the number fluctuations are known as most relevant sources of multifractality in series. Typical way to clarify this question is to analysis the corresponding shuffled and surrogate series. The former destroys all types of correlation while the latter corresponds to Gaussianity of probability density function, accordingly the PDF is altered to Gaussian function.  
         In the case of multifractality dominated by contribution of various correlations in data, considerably, therefore the scaling exponent of fluctuation function is more affected by shuffling procedure. While, the multifractality nature due to the fatness of the PDF signals almost remains unchanged by the shuffling routine. We compared the generalized Hurst exponent, $h(q)$ , for the original series with the result of the corresponding shuffled series. Difference between these two functions directly indicates the presence of correlations in the original series. In Fig. \ref{figpdf} we indicate the probability density function of series. This plot confirms that departure from Gaussian distribution (thick solid line in Fig. \ref{figpdf}) can hardly be recognized. 
         
\begin{figure}
\begin{center}
\includegraphics[width=1\linewidth]{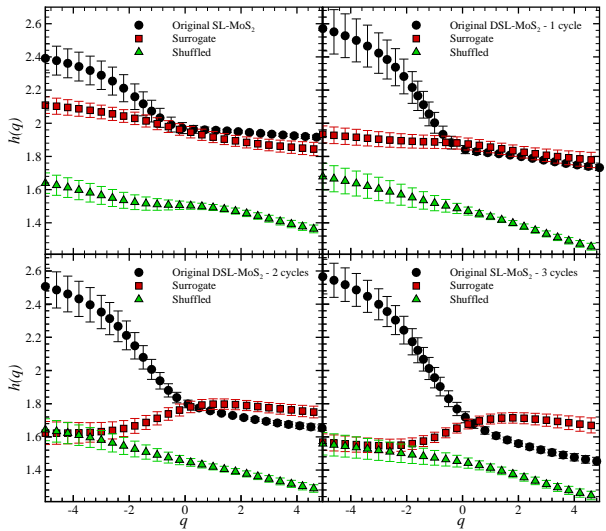}
\caption{\label{fig7} Generalized Hurst exponent, $h(q)$, for original sample, surrogate and shuffled. Upper left panel corresponds to SL-${\rm MoS_2}$. Upper right panel indicate DSL-${\rm MoS_2}$-1 cycle. Lower left panel illustrates DSL-${\rm MoS_2}$-2 cycles and  lower right panel shows DSL-${\rm MoS_2}$-3 cycles.}
\end{center}
\end{figure}

\begin{figure}
\begin{center}
\includegraphics[width=1\linewidth]{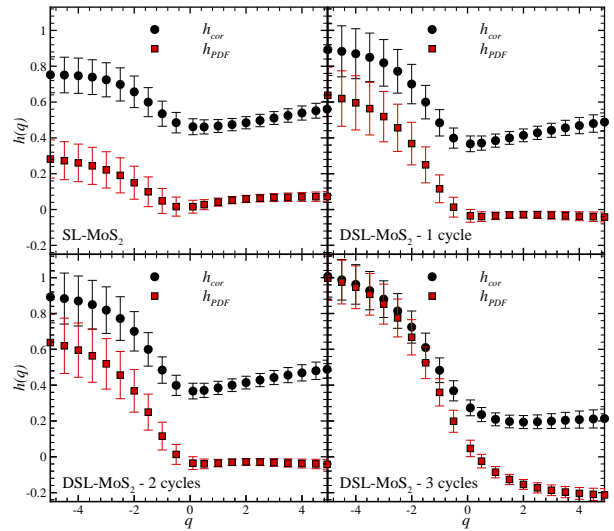}
\caption{\label{fig8} $h_{\rm PDF}(q)$ and $h_{\rm{cor}}(q)$ for SL-${\rm MoS_2}$ (upper left panel), DSL-${\rm MoS_2}$-1 cycle (upper right panel), DSL-${\rm MoS_2}$-2 cycles (lower left panel) and DSL-${\rm MoS_2}$-3 cycles (lower right panel).}
\end{center}
\end{figure}

The generalized Hurst for original, shuffled and surrogated time series are shown in Fig. \ref{fig7}. The $q$-dependency of $h_{\rm PDF}(q)=h(q)-h_{\rm surrogate}(q)$ and $h_{\rm cor}(q)=h(q)-h_{\rm shuffled}(q)$ demonstrate that the multifractality of the PL spectra is due to both broadness of the PDF and long-range correlation \cite{kant02,shan08}. Since, if the multifractality only belongs to the long-range correlation, we should have $h_{\rm shuffled}(q)=0.5$. As shown in Fig. \ref{fig8}, the value of $h_{\rm cor}(q)$ is greater than $h_{\rm PDF}(q)$, so the multifractality due to the fatness is weaker than the multifractality due to the correlation. As discussed before, for $q>0$ the large fluctuations have more contribution in determining $h(q)$, while for $q<0$, the small fluctuations play dominant role. In our analysis, we found that increasing amount of defect destroys existence correlation in the intensity increment ($\delta I$). Subsequently, we expect to have smaller value for $h_{\rm cor}(q=2)$ for DSL-${\rm MoS_2}$-3 cycles comparing to SL-${\rm MoS_2}$. This behavior remains almost for other positive value of $q$. On the contrary by increasing defects in our sample, it seems that small fluctuations are statistically grow up leading to have opposite trend. But our results for $q<0$ have large error-bars and we can not recognize a significance difference between our samples in this regime. 

Final remark is that photoluminescence occurs when a semiconductor is excited by photon (here, green light). Accordingly, the electrons absorb energy and excited state returns to the ground state by emission of radiation. There are several possibilities of returning to the ground state. The observed emission from a luminescent media such as SL-${\rm MoS_2}$ is a process of returning to the ground state radiatively.  The processes competing with luminescence are radiative transfer to another ion and nonradiative transfers such as multiphonon relaxation and energy transfer between different ions in semiconducting crystal.  In perfect semiconductors, there exist a band gap (forbidden band) composed of valence and conduction band. Point defects such as S-vacancy in SL-${\rm MoS_2}$ with functionality same as dopants act as donors or acceptors; donors donate an electron to the conduction band, whereas acceptors accept an electron from the valence band. When defects such as impurities and vacancies are created, some allowed energy states are imposed somewhere in the band gap. The states that are close to the band edges (either conduction or valence band), called shallow traps, and the states are close to the middle of band gap corresponding to deep traps produce shallow and deep energy levels.  Various radiative transitions between the original and defect-induced energy levels leads to the fluctuations in PL spectrum. Inspired by multifractal properties of self-similar phenomena, it is interesting to utilize multifractal analysis of PL spectrum fluctuations in order to distinguish between different types of maters according to different observables. Stochasticity in recorded fluctuations enforce to utilize robust statistical analysis. Therefore, multifractal detrended analysis methods are proper tools in this regard.

\section{Conclusion}

     Semiconducting 2D materials, such as single layer SL-${\rm MoS_2}$, have the intrinsic large band gap and flexible structure to apply in nano/optoelectronic devices, transistors, logic circuits, sensors and photocatalysis. Almost these applications need single or few layer without high population of point defects such as sulfur vacancies created during synthesis stages. Though photoluminescence (PL) spectroscopy is a common analyzing method to identify the band gap and layer number but it has never been utilized to reveal presence of defects. To the best of our knowledge, for the first time we carried out multifractal detrended fluctuation analysis (MF-DFA) and multifractal detrended moving average (MF-DMA) to extract proper information from PL spectrum to determine effect of defects on PL spectra of SL-${\rm MoS_2}$.  MF-DFA and MF-DMA have been shown to be the efficient nonlinear analytical techniques to provide quantitative estimates of stochasticity in the nonlinear and non-stationary data. Two parameters related to multifractality including generalized Hurst exponents, $h(q)$, and multifractal singularity spectra reveal that 1) PL spectrum of SL-${\rm MoS_2}$ representing photon- inducing electron transition from valence band to conduction band has multifractal nature. 2) The presence of the point defects on SL-${\rm MoS_2}$ alters the PL spectrum so the multifractality characteristics would be dramatically changed. 3) Smaller fluctuation in emitted light can be obtained by increasing lattice defects. So that not only quality of SL-${\rm MoS_2}$ can be identified by studying multifractality of PL spectra but also it possesses estimation about relative defect populations without any more expensive instruments such as high-resolution transmission electron microscopy.
It is interesting to extend this study for other materials. In addition, including the  preprocesses such as adaptive detrending (AD) \cite{jhu09} and singular value decomposing (SVD) \cite{nag05} can be used for same purpose. Cross-correlation between different measurements such as PL and Raman spectroscopies, in order to get deep insight for characterizing Graphene-like materials, are other researches based on multifractal detrended fluctuation cross-correlation analysis (MF-DXA) \cite{zhou08}. We will apply them for other studies.  

{\bf Acknowledgements}  Thanks the anonymous referee for help us to improve this paper. SMSM is grateful to school of Physics, Institute for research in fundamental sciences (IPM), where some parts of this paper have been finalized.

%Woo Cheol Jun, Gabjin Oh and Seunghwan Kim, PHYSICAL REVIEW E 73, 066128 (2006), Understanding volatility correlation behavior with a magnitude cross-correlation function

%Long-range cross correlations between two Stocks imply that each stock separately has long memory of its own previous values and, additionally, has a long memory of previous values of the other stock (from podobnik)

%\begin{acknowledgements}
%\end{acknowledgements}



\end{document}